\pdfoutput=1

\documentclass[11pt]{article}

\usepackage[final]{acl}

\usepackage{times}
\usepackage{latexsym}

\usepackage[T1]{fontenc}

\usepackage[utf8]{inputenc}

\usepackage{microtype}

\usepackage{inconsolata}

\usepackage{graphicx}
\usepackage{amsfonts}
\usepackage{array}
\usepackage{booktabs}
\usepackage{amsmath}
\usepackage{cleveref}
\usepackage{enumitem}
\title{A Diverse and Effective Retrieval-Based Debt Collection System with Expert Knowledge}

\author{Jiaming Luo\textsuperscript{1}, Weiyi Luo\textsuperscript{2}, Guoqing Sun\textsuperscript{2}, Mengchen Zhu\textsuperscript{2}, Haifeng Tang\textsuperscript{2},\\
\textbf{Kunyao Lan\textsuperscript{1}, Mengyue Wu\textsuperscript{1}\footnotemark[1], Kenny Q. Zhu\textsuperscript{3}\thanks{Corresponding authors.}}\\
\textsuperscript{1} 
X-LANCE Lab, Department of Computer Science and Engineering \\
 MoE Key Lab of Artificial Intelligence, AI Institute \\
  Shanghai Jiao Tong University, Shanghai, China \\
\textsuperscript{2}China Merchants Bank Credit Card Center, Shanghai, China  \\
\textsuperscript{3}University of Texas at Arlington, Arlington, Texas, USA \\
\texttt{\textsuperscript{1}\{leojm2017, lankunyao, mengyuewu\}@sjtu.edu.cn},\\
\texttt{\textsuperscript{2}\{luoweiyi, gqsun, zmc1996, thfeng\}@cmbchina.com},\\
\texttt{\textsuperscript{3}kenny.zhu@uta.edu}
}

\begin{document}
\maketitle
\begin{abstract}

Designing effective debt collection systems is crucial for improving operational efficiency and reducing costs in the financial industry. However, the challenges of maintaining script diversity, contextual relevance, and coherence make this task particularly difficult. This paper presents a debt collection system based on real debtor-collector data from a major commercial bank. We construct a script library from real-world debt collection conversations, and propose a two-stage retrieval based response system for contextual relevance. Experimental results show that our system improves script diversity, enhances response relevance, and achieves practical deployment efficiency through knowledge distillation. This work offers a scalable and automated solution, providing valuable insights for advancing debt collection practices in real-world applications.

\end{abstract}

\section{Introduction}

Debt collection plays a crucial role in the financial industry. In practice, outbound calls for debt recovery are typically handled by experienced experts, since negotiating with debtors is often challenging. Consequently, large companies must employ substantial number of staff to manage daily debt collection tasks, leading to high operational costs. This has spurred interest in developing systems that assist human experts or automate outbound calls, making it a burgeoning area of research \cite{zhang-etal-2018-modeling,wang2020two}.

\begin{figure}[ht]
    \centering
    \includegraphics[width=\columnwidth]{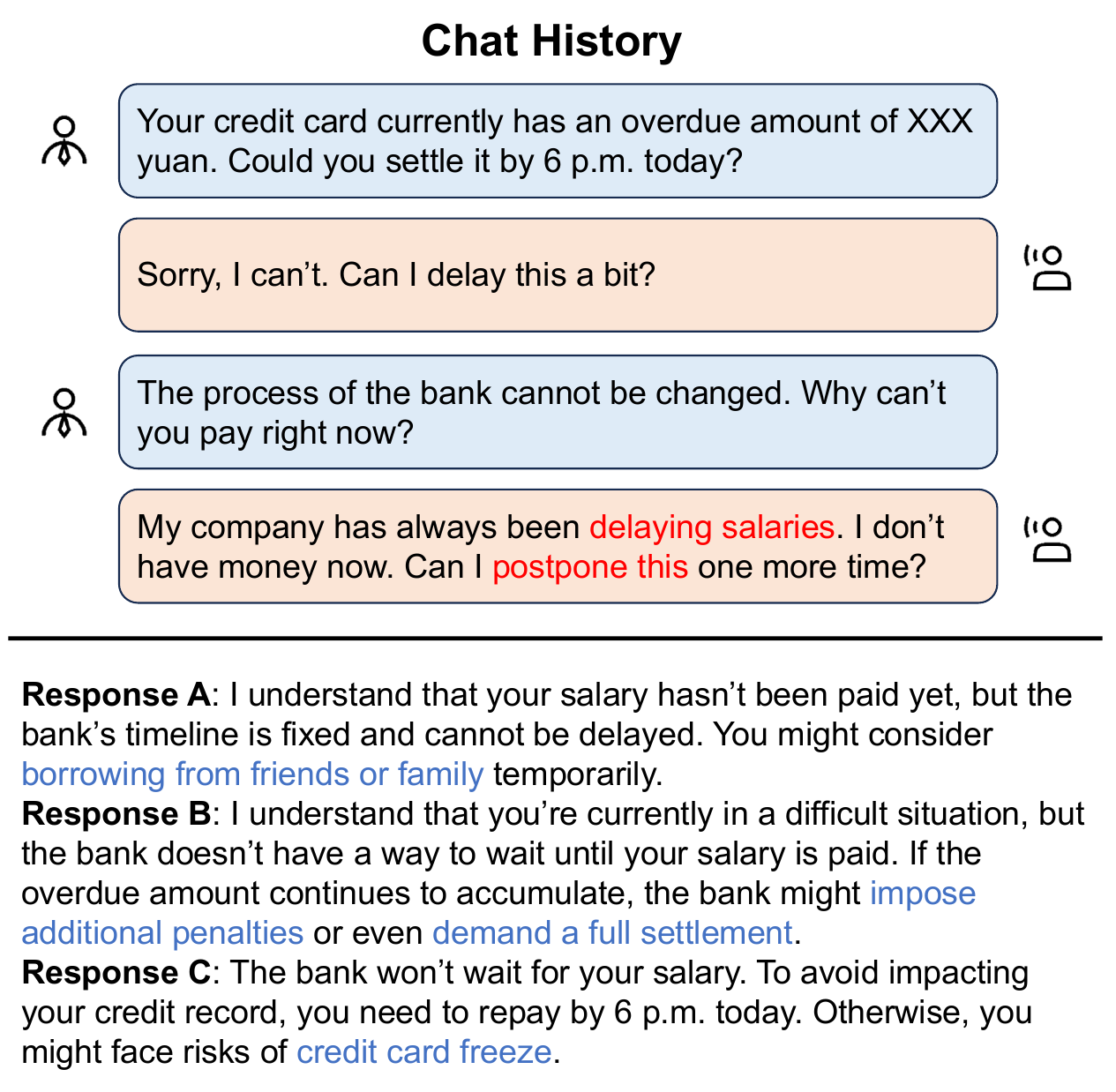} 
    \caption{An exemplar between a debtor and a collector, with three candidate responses. The debtor's intent is labeled in red while the strategies in collector's responses are labeled in blue.}
    \label{fig:conversation}
\end{figure}

Recent advancements have demonstrated the feasibility of automatic outbound agents \cite{zhang2023towards, wang2020two}. Currently, many collection chatbots are flow-based systems configured with rule-based frameworks authored by experts \cite{wang2020two,jia2020matching}. In these systems, the chatbot predicts the debtor's intent at each stage and provides predefined responses based on established rules. However, such flow-based systems face notable limitations. They heavily depend on expert-crafted rules, making them difficult to update and scale to different scenarios due to their complexity. Additionally, these systems lack response diversity, as the output is fixed for each scenario.

To address these limitations, researchers have explored using pretrained language models to generate responses based on dialogue context \cite{zhang2023towards,jin2023joint,jia2020multi,zhang-etal-2023-semantic}. These methods eliminate the need for predefined rules by fine-tuning models on large-scale debt collection conversations. However, generative models often produce responses that may be ineffective in debt collection. The responses are also difficult to control due to their inherent uncertainty.

In view of these problems, retrieval based response system become a better choice in practice, as the response outputs are more controllable. Typically, it consists of two stages: script\footnote{The term ``script'' refers to predefined response or standardized dialogue templates used by debt collection agents.} generation and response system implementation. As for script generation, current practice remains predominantly a manual process undertaken by experienced experts. However, previously-mentioned challenges still exist. First, achieving script diversity is inherently challenging, as generating distinct responses for a wide range of scenarios demands significant effort. Second, updating the system is resource-intensive, requiring expert intervention to craft and integrate new scripts with each revision. To address these issues, automatic script generation from real conversations has become a promising direction. 

On the other hand, response retrieval in debt collection is particularly challenging due to several factors. In practice, we find that embedding-based methods, while effective in other domains \cite{su2023dial,zhang2022two}, struggle here due to the difficulty of distinguishing positive from negative samples without manual annotation. Typically, positive samples are selected from actual responses in the dialogue, and negative samples are randomly chosen from other dialogues. However, this random negative sampling often leads to situations where the selected ``negative'' samples are actually suitable responses for the current dialogue, resulting in ``false negatives''. 

This increases the complexity of model training and affects the accuracy of the system, particularly when multiple responses in the script library appear valid during inference.
To this end, we propose a two-stage retrieval based response system to select the most effective script from script library.

\begin{figure}[ht]
    \centering
    \includegraphics[width=\columnwidth]{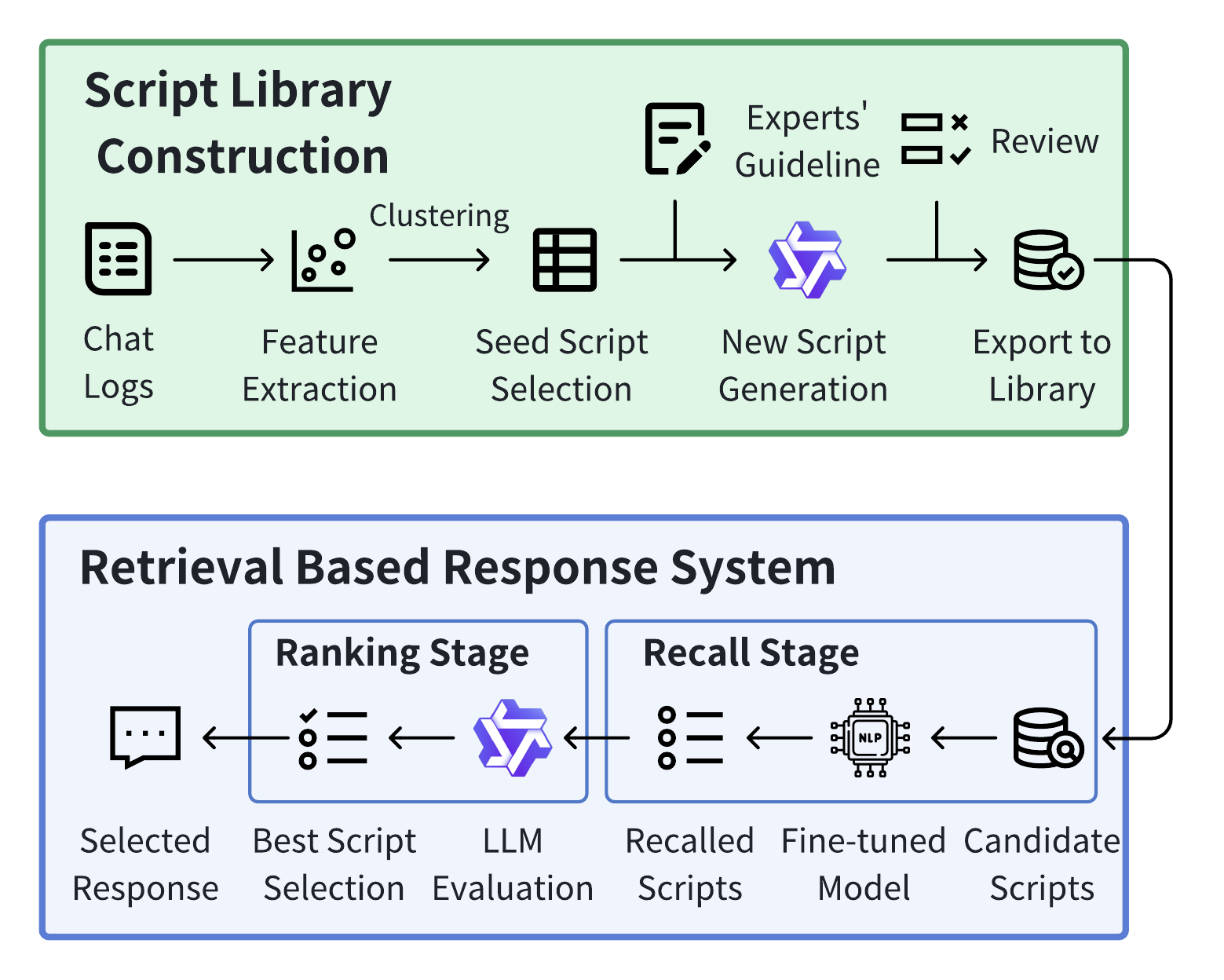} 
    \caption{Overview of the SCORES framework. A script library is constructed from chat logs, followed by a two-stage response selection system.}
    \label{fig:main_figure2}
\end{figure}

In this work, we propose a comprehensive system for automatic outbound chat-bots that integrates script generation and selection models. Leveraging the capabilities of Large Language Models (LLMs), we first generate diverse and effective scripts based on real-world conversations while incorporating expert knowledge to enhance the quality and naturalness of the dialogues. After that, to ensure the safety and appropriateness of outbound calls, we frame the problem as response selection, where the model must choose the optimal response given the dialogue context. Since traditional embedding-based models often struggle to distinguish between similar scripts with identical strategies, to overcome this, we design a two-stage retrieval pipeline: the first stage employs a pretrained model to recall $n$ relevant responses, and the second stage uses LLMs to evaluate and select the best response based on three aspects: \textit{Empathetic Engagement}, \textit{Effective Problem-Solving} and \textit{Contextual Relevance}. The contributions of our work are threefold: 
\begin{enumerate}[topsep=0pt, partopsep=0pt, itemsep=0pt, left=0pt]
   \item A novel framework to generate high-quality scripts by leveraging insights from human conversations and domain expertise, where we automatically obtain more than 1,000 scripts for 9 strategies accepted by experts.
   \item A two-stage retrieval pipeline that efficiently tackles the challenges of response selection, elevating Recall@1 from 0.346 to 0.577.
   \item An automatic outbound framework SCORES: \textbf{S}cript \textbf{C}reation and \textbf{O}ptimized \textbf{RE}sponse \textbf{S}ystem, a scalable and practical pipeline with minimal supervision that can easily extend to other related domains including marketing and intelligent customer service. 
\end{enumerate}

\section{Methodology}

Our proposed framework SCORES includes two modules: (a) Automatic script library construction and (b) Retrieval based collection system. 

\subsection{Automatic Script Library Construction}
The script library is a fundamental component of the debt collection system. Our approach utilizes LLMs to generate diverse scripts based on real debt collection dialogue history from a major bank's Debt Recovery Department that involves a large amount of daily debt collection calls.

\paragraph{Data Preparation}
We begin by collecting voice recordings of interactions between debtors and collectors over several days from a major commercial bank. These recordings are transcribed using an automatic speech recognition (ASR) system. Each dialogue transcription $T$ between a debtor $D$ and a collector $C$ is organized into a turn-taking format: $T=\{c_1,d_1,c_2,d_2,\dots,c_n,d_n\}$, where $c_i$ and $d_i$ represent the collector's and debtor's utterances, respectively. Each collector's utterance $c_i$ is assigned a strategy label $s_i\in S$, (e.g., ``Pressure through letters'', ``Pressure through family''), and each debtor's utterance $d_i$ is assigned with purpose label $p_i\in P$, (e.g., ``Inability to repay'', ``Unemployment''). Here $S$ is the pre-defined strategy list while $P$ is the pre-defined purpose list. These labels can be annotated by experts or automatically extracted using fine-tuned language models.
Next, we extract utterance pairs $[d_i,c_i]$ from each dialogue and filter out pairs without applicable strategy or purpose labels. This process results in a collection of $m$ labeled utterance pairs $U = \{d_i,c_i\}, i\in \{1,\dots,m\}$.

\paragraph{Seed Scripts Selection}
Everyday conversation data contains diverse debt collection strategies. However, variations in speaking styles and scenarios make it challenging to generalize patterns for each strategy. To address this, we select seed scripts for each strategy from the utterance pairs $U$. 
We first divide collectors' utterances by strategy and use embedding models to represent each utterance as a $d$ dimensional vector $e_i$. Here we employ BGE-M3 \cite{chen2024bge} to extract 1024-dimensional embeddings. These embeddings are clustered using the K-means algorithm, producing $K$ clusters: $\mathbb{E} = \{E_1,E_2,\dots,E_K\}, E_i = \{e_1^i,e_2^i,\dots,e_j^i\}$. The mean of each cluster's embeddings is computed as the cluster center: $o_i = \frac{1}{j}\sum_{m=1}^{j}e_m^i$.
For each cluster, we select the top-5 embeddings closest to the center as representative ``seed scripts''. These scripts capture distinct ``persuasive patterns'' for the strategy. This process yields $5\times K$ seed scripts for each strategy.

\paragraph{Script Generation}
Using the selected seed scripts, we generate additional scripts tailored for debt collection using Qwen2-72B~\cite{yang2024qwen2technicalreport}.
To ensure contextual fluency and coherence, generated scripts are aligned with the debtor's purpose $p_i$. We incorporate expert guideline for each purpose into the generation process. For example, if a debtor mentions its unemployment during the conversation, the response should first empathize and then proceed with the standard collection strategy. In practice, the purpose-specific guidelines and the seed scripts are input into the LLM to generate three new scripts per cluster. These scripts are labeled with $p_i$ and $s_i$ for subsequent use. 
Generated scripts are reviewed and refined by experts before being added to the script library, whose results are illustrated in \Cref{sec:results}.

\subsection{Retrieval-based Response System}
The response system generates or retrieves responses during debt collection conversations. We adopt a retrieval-based approach for safety and reliability. Our response selection pipeline consists of two stages: \textit{recall} and \textit{ranking}. The recall stage is designed to efficiently narrow down a large pool of candidate responses to a smaller subset that is contextually relevant to the conversation. The ranking stage then refines this subset, selecting the most appropriate response based on LLM evaluations. This two-stage process ensures both scalability in handling a large response database and precision in selecting high-quality responses.

\paragraph{Recall Stage}
The recall stage identifies the top-$n$ candidate scripts from the library. Given a context history $h_i$ and the purpose $p_i$ of the debtor's last utterance, the recall model retrieves the most appropriate scripts labeled with $p_i$.
We pre-process conversation transcriptions by dividing them into sub-conversations using a sliding window. Each sub-conversation consists of five utterances as context $h_i$ and the sixth utterance as the response $r_i$: $h_i = \{d_i,c_{i+1},d_{i+1},c_{i+2},d_{i+2}\}, r_i = c_{i+3}$

Following prior work on response selection \cite{su2023dial}, we use Chinese-BERT-wwm \cite{cui-etal-2021-pretrain} as the base model $M$. The model is first pretrained with a Masked Language Modeling (MLM) objective and fine-tuned using contrastive learning:

\begin{equation}
    \mathcal{L} = \sum_{i=1}^m \frac{\exp(w_i^+)}{\exp(w_i^+) + \sum_{j=1}^{n_{neg}} \exp(w_i^j)}
\end{equation}
where $w_i^+ = \text{sim}(h_i, r_i^+), w_i^j = \text{sim}(h_i, r_i^{j-})$. $r_i^+$ is the correct response, $r_i^{j-}$ are negative samples, and $\text{sim}(h_i,r_i)$ is the cosine similarity between embeddings. We use the [CLS] token of the last hidden layer of $M$ as the embedding of the texts. 

After fine-tuning, the model $M$ encodes $h_i$ and candidate scripts into embeddings. During inference, $M$ generates embeddings for the given context, and the top-$n$ most similar scripts are retrieved as recall results.

\paragraph{Ranking Stage}
Although the recall stage reduces the pool of candidate responses, selecting the best script remains challenging due to the nuanced, indirect alignment between the conversational context and the desired strategy. To address these issues, we leverage LLMs to evaluate and select the best response from the candidates chosen in the recall stage. An intuitive approach involves assessing candidate scripts based on several predefined aspects. After consulting with debt collection experts, we identified three critical aspects for evaluation: \textit{Empathetic Engagement}, \textit{Effective Problem-Solving} and \textit{Contextual Relevance}. Detailed definitions can be found in \cref{sec:aspect}.

Inspired by G-Eval \cite{liu2023g}, we define three levels for each aspect: excellent (3), good (2), and poor (1). Each level is supported by detailed criteria, crafted by experts. During the evaluation process, we combine the context in 3 turns and each candidate script into a prompt template, instructing the LLM to score the script according to the predefined criteria (see \cref{sec:prompt}). The average score across the three aspects serves as the overall score for each candidate. The script with the highest overall score is selected as the response. In cases of tied scores, the script ranked higher in the recall stage is chosen.

Despite the effectiveness of LLM evaluation, the inference time for large models, such as Qwen2-72B, is prohibitively high for real-time response systems. To mitigate this, we apply a knowledge distillation approach, transferring expertise from the large LLM (72B-model) to a more computationally efficient small LLM (1.5B/3B-model). Specifically, we use Qwen2-72B model to generate labeled data by evaluating context-candidate pairs using the predefined criteria. These evaluation scores and accompanying rationales serve as the labels.

We then fine-tune smaller LLMs (e.g., Qwen2.5-3B \cite{qwen2.5}) on the labeled dataset. We set the context-candidate pair and evaluation criteria as inputs, while the evaluations generated by the Qwen2-72B model are the desired outputs. After fine-tuning, the smaller LLM can efficiently perform ranking, significantly reducing inference time while maintaining acceptable performance. For example, the Recall@1 metric for the Qwen2.5-3B model improved significantly from 0.404 to 0.577 after fine-tuning. Additional experimental results are provided in \Cref{sec:rank-results}.

\section{Experiments}

In this section, we present the experimental settings and results of our proposed methods. 

\subsection{Datasets}

For script library construction, we processed 786 debt collection calls, transcribed them using ASR tools, and annotated debtor utterances with a pretrained purpose classification model. LLMs identified collector strategies, and experts refined the annotations, yielding 6,218 labeled utterances. All our data is in Chinese.

For response system construction, we transcribed 4,000 additional calls and used the classification model to annotate purposes without further human review. After segmenting dialogues, we obtained over 40,000 context-response pairs, split into training, validation, and test sets (8:1:1). For knowledge distillation, the Qwen2-72B model generated 13,000 cases in Alpaca format.

\subsection{Implementation Details}

\paragraph{Script Library Construction} We used the BGE-M3 model to encode sentences into 1024-dimensional vectors. For seed script selection, the utterances were clustered into $K=4$ groups using K-means, and five utterances nearest to each cluster center were selected as seed scripts. We use Qwen2-72B model for script generation.

\paragraph{Response System Construction} 
We used the Chinese-BERT-wwm model with a truncation length of 256. Pretraining employed a 30\% masking ratio, a $1\times10^{-4}$ learning rate, and five epochs, selecting the best model via validation. Fine-tuning used a $5\times10^{-5}$ learning rate, a batch size of 64, and AdamW \cite{loshchilov2019adamw} optimizer for five epochs. For ranking, Qwen2.5-3B and Qwen2.5-1.5B models were fine-tuned with LoRA \cite{hu2021lora} on the LLaMA-Factory platform \cite{zheng2024llamafactory}. All experiments ran on a V100 GPU server.

\subsection{Results and Discussions}
\label{sec:results}

\paragraph{Script Library} 
To evaluate the effectiveness of K-means clustering, we calculated the intra-cluster distance $d_{\text{intra}}$ and the inter-cluster distance $d_{\text{inter}}$. For clusters corresponding to each strategy, $\mathbb{E} = \{E_1, E_2, \dots, E_K\}$, where $E_i = \{e_1^i, e_2^i, \dots, e_j^i\}$, the intra-cluster distance for each cluster is computed as the average distance between all embeddings and the cluster center:

\begin{equation}
d_{\text{intra}} = \frac{1}{K}\frac{1}{|E_i|}\sum_{i=1}^{K} \sum_{k=1}^{|E_i|} d(x_k^i, o_i)
\end{equation}
Here, $o_i$ denotes the center of cluster $i$, and $d(x, y)$ represents the L2 distance between two vectors.
Similarly, the inter-cluster distance is calculated as the average distance between embeddings within a cluster and the centers of all other clusters:

\begin{equation}
d_{\text{inter}} = \frac{1}{K} \frac{1}{K-1} \frac{1}{|E_i|} \sum_{i=1}^{K} \sum_{j \neq i}^{K} \sum_{k=1}^{|E_i|} d(x_k^i, o_j)
\end{equation}
These metrics assess the compactness of clusters and the separability between different strategies.

From \Cref{distance}, we observe that the intra-cluster distance is smaller than the inter-cluster distance, which demonstrates the effectiveness of the clustering method. This result indicates that seed scripts within the same cluster exhibit higher similarity (consistency), while those across different clusters show greater variation (diversity).

\begin{table}[]
    \centering
    \small
    \caption{Intra-distance and inter-distance comparison.} \label{distance}
    \[
    \begin{array}{lcc}
    
    \toprule
    \textbf{Strategy} & \textbf{$d_{\text{intra}}$} & \textbf{$d_{\text{inter}}$} \\
    \midrule
    \text{Pressure Through Letters} & 0.3361 & 0.4910 \\
    \text{Card Suspension} & 0.2921 & 0.5144 \\
    \text{Full Payment}  & 0.3126 & 0.4743 \\
    \text{Negotiation Plan} & 0.3306 & 0.4984 \\
    \text{Cash Advance} & 0.4382 & 0.5178 \\
    \text{Pressure Through Family} & 0.3613 & 0.6021 \\
    \text{Credit Report} & 0.3363 & 0.4708 \\
    \text{Repayment Ability} & 0.3959 & 0.4683 \\
    \text{Anti-Disconnection} & 0.3116 & 0.4992 \\
    \midrule
    \textbf{Average} & \textbf{0.3491} & \textbf{0.4946} \\
    \bottomrule
    \end{array}
    \]
    
\end{table}

To further assess the diversity of generated scripts, we compute the Distinct-n metrics \cite{li-etal-2016-diversity} under different seed script selection methods. Random refers to selecting 5 utterances randomly as seed scripts for each strategy. The configurations $k=1$ and $k=4$ differ in the number of clusters. Specifically, $k=1$ means selecting the top-5 utterances closest to the center of all strategy embeddings, whereas $k=4$ involves clustering the utterances into four groups and selecting 5 utterances nearest to the center of each cluster.

\begin{table}[t]
\centering
\small
\caption{Distinct-n evaluation across different seed script selection strategies. The best results are in bold.} \label{distinct_table}
\[
\begin{array}{lcc}
\toprule
\textbf{Selection} & \textbf{Distinct-1} & \textbf{Distinct-2} \\
\midrule
Random & 0.131 & 0.466 \\
k=1 & 0.129 & 0.466 \\
k=4 & \textbf{0.141} & \textbf{0.500} \\
\midrule
\end{array}
\]
\end{table}

We evaluate the diversity using scripts generated by the same LLM across 5 randomly sampled purposes and 9 predefined strategies (as listed in \Cref{distance}). The total number of generated scripts for the Random and $k=1$ settings is $5 \times 9 \times 3 = 135$. For the $k=4$ setting, we generate $3 \times 4 = 12$ scripts for each purpose-strategy pair and randomly sample 3 scripts, maintaining the evaluation size at 135 scripts for comparability. We evaluate the diversity using Distinct-1 and Distinct-2, where higher scores indicate greater diversity.

As shown in \Cref{distinct_table}, the $k=4$ configuration achieves the highest Distinct-n scores among the three settings. This result demonstrates that the clustering-based method effectively generates scripts that are both diverse and consistent within their respective clusters.

We further evaluate the script library's performance in real-world scenarios. For an A/B test, we replaced the existing scripts with those generated by the LLM while keeping the chatbot workflow unchanged. During a month-long online test involving approximately 600,000 outbound calls, the script replacement led to a 0.5\% improvement in recovery rate. This shows the effectiveness of our script generation method.

\paragraph{Recall Stage}
We evaluate the performance of our fine-tuned model in the recall stage using Recall@K (R@K) on the test set. In this evaluation, the candidate set contains 10 utterances, in which 1 utterance is designated as the ground truth. We compare the model's performance with and without the pretraining stage. As shown in \Cref{tab:pretrain_comparison}, the model's performance improves significantly with the inclusion of the pretraining stage.

\begin{table}[h]
\centering
\small
\caption{Performance comparison w/ or w/o pretraining}
\label{tab:pretrain_comparison}
\[
\begin{array}{lcccc}
\toprule
\textbf{Model} & \textbf{R@1} & \textbf{R@2} & \textbf{R@3} & \textbf{R@5} \\
\midrule
\text{w/~ pre.}  & \textbf{0.617} & \textbf{0.782} & \textbf{0.870} & \textbf{0.957} \\
\text{w/o~ pre.} & 0.594 & 0.762 & 0.859 & 0.951 \\
\bottomrule
\end{array}
\]
\end{table}

When comparing these results to those reported in the E-Commerce Dataset \cite{su2023dial}, the Recall@K metrics are noticeably lower. For example, R@1 for the baseline model (BERT+CL) reaches 0.849 in the E-Commerce dataset but only achieves 0.671 in our dataset. This highlights the complexity of our response selection task, underscoring the necessity of adopting a two-stage selection pipeline to address these challenges effectively.

\paragraph{Ranking Stage}
\label{sec:rank-results}
To evaluate the performance of different models in the ranking stage, we employed 7 debt collection experts to select the best response for a given context from 3 candidate utterances from the recall stage. The most frequently selected utterance is regarded as the ground truth. In total, 52 cases were labeled as the test set, with a Fleiss' kappa value of 0.41, indicating ``Moderate Agreement.'' This highlights the inherent difficulty of selecting the best response from candidates from the recall stage.

\begin{table}[h]
\centering
\caption{Performance comparison of ranking models on Recall@1. Models with the ``-sft'' suffix denote the models are supervised fine-tuned on the dataset labeled by Qwen2-72B. ``BERT'' refers to the fine-tuned model used in the recall stage, while ``72B'' represents Qwen2-72B, ``3B'' represents Qwen2.5-3B, and ``1.5B'' represents Qwen2.5-1.5B. }
\label{tab:ranking}
\resizebox{\linewidth}{!}{
\begin{tabular}{lcccccc}
\toprule
\textbf{Model} & \textbf{BERT} & \textbf{72B} & \textbf{3B-sft} & \textbf{3B} & \textbf{1.5B-sft} & \textbf{1.5B} \\
\midrule
\textbf{R@1}  & 0.346 & \textbf{0.731} & 0.577 & 0.404 & 0.538 & 0.423 \\
\bottomrule
\end{tabular}
}
\end{table}

Then we compared the performance of 5 LLMs against the BERT model baseline by evaluating Recall@1 on the labeled test set. The results are summarized in \Cref{tab:ranking}. The results indicate that the performance of the recall stage remains suboptimal, with Recall@1 slightly surpassing random guess (0.333). Despite this, score-based methods using LLMs demonstrate promising results. Notably, the 72B-model, even without supervised fine-tuning, shows a significant improvement over the baseline. Similarly, the 3B and 1.5B models also outperform the baseline, highlighting the potential of LLMs as effective ranking models for complex tasks.

Moreover, after distilling knowledge from the 72B model, the performance of the 3B and 1.5B models improves significantly. This demonstrates the feasibility of leveraging smaller LLMs in real-world applications by distilling knowledge from larger models.

\section{Related Work}

\paragraph{Retrieval-Based Dialogue Systems}
Retrieval-based dialogue systems aim to identify the most appropriate response from a set of candidates \cite{jia2021ddrel,jin2023joint}. These systems are widely applied in domains such as customer service Q\&A and forum post interactions \cite{lowe2015ubuntu,zhang-etal-2018-modeling,wu2016sequential}. Modern approaches predominantly leverage pre-trained language models (PLMs) like BERT \cite{devlin2019bert}, fine-tuned using contrastive learning on domain-specific corpora \cite{xu2021learning,zhang2022two,zhang-etal-2023-semantic}. To enhance semantic relevance and contextual coherence, Han et al. \cite{han2021fine} incorporate fine-grained labels during post-training. Su et al. \cite{su2023dial} propose a novel post-training method that improves context embeddings. Additionally, Han et al. \cite{han-etal-2024-efficient} introduce EDHNS, which optimizes contrastive learning by focusing on harder-to-distinguish negative examples.

\paragraph{Automatic Outbound Chatbots}
Automatic outbound chatbots are designed to engage customers in conversations to achieve specific goals, such as debt collection or advertising. Traditional systems often relied on flow-based approaches due to their straightforward logic and ease of implementation \cite{lee2008implementation,yan2017building}. However, these systems heavily depend on expert-defined rules and are challenging to update. To address these limitations, recent research has shifted towards response generation using PLMs. Jin et al. \cite{jin2023joint} propose a persuasion framework that integrates both semantic understanding and strategic considerations. Zhang et al. \cite{zhang2023towards} enhance response generation by incorporating user profiles extracted during conversations. Qian et al. \cite{qian2022toward} redefine the dialogue process as a sequence-labeling problem, leveraging a dual-path model for joint multi-task learning.

\section{Conclusion}

In this work, we designed and evaluated a comprehensive system, SCORES, for automating outbound debt collection, addressing challenges of script diversity, adaptability, and effective response selection. By combining the script generation capabilities of LLMs with a robust two-stage retrieval framework, we achieved notable improvements in response effectiveness. Besides, knowledge distillation enhanced its efficiency for real-world deployment. More importantly, the flexibility of this framework allows it to be adapted to a wide range of domains, such as customer support and telemarketing. Future work will focus on further refining script diversity, improving real-time response evaluation, and expanding the framework’s applicability to ensure even higher levels of performance and adaptability in diverse settings.

\section*{Ethical Considerations}
In our experiments, call records were collected with customer consent. To ensure data privacy, personal information such as names and phone numbers was removed during script generation and further training. When testing online, the responses generated by SCORES are exclusively retrieved from the script library, where all scripts were carefully reviewed to eliminate any inappropriate content.

\section*{Acknowledgements}
This work has been supported by the CMB
Credit Card Center \& SJTU joint research grant and Guangxi major science and technology project~(No. AA23062062).

\bibliography{acl_latex}

\appendix

\section{Appendix}
\label{sec:appendix}

\subsection{Evaluation Aspects for LLM}
\label{sec:aspect}
\begin{enumerate}[topsep=0pt, partopsep=0pt, itemsep=0pt, left=0pt, labelsep=0pt]
    \item \textbf{Empathetic Engagement:} This aspect evaluates the politeness and the ability to show empathy and understanding for the debtor's difficulties.
    \item \textbf{Effective Problem-Solving:} This aspect assesses whether the script effectively communicates the consequences of contract breaches and provides a viable solution.
    \item \textbf{Contextual Relevance:} This aspect determines whether the script maintains logical coherence with the preceding text.
\end{enumerate}

\subsection{Prompt for LLM Evaluation}
\label{sec:prompt}

\begin{figure*}[ht]
    \centering
    \includegraphics[width=\textwidth]{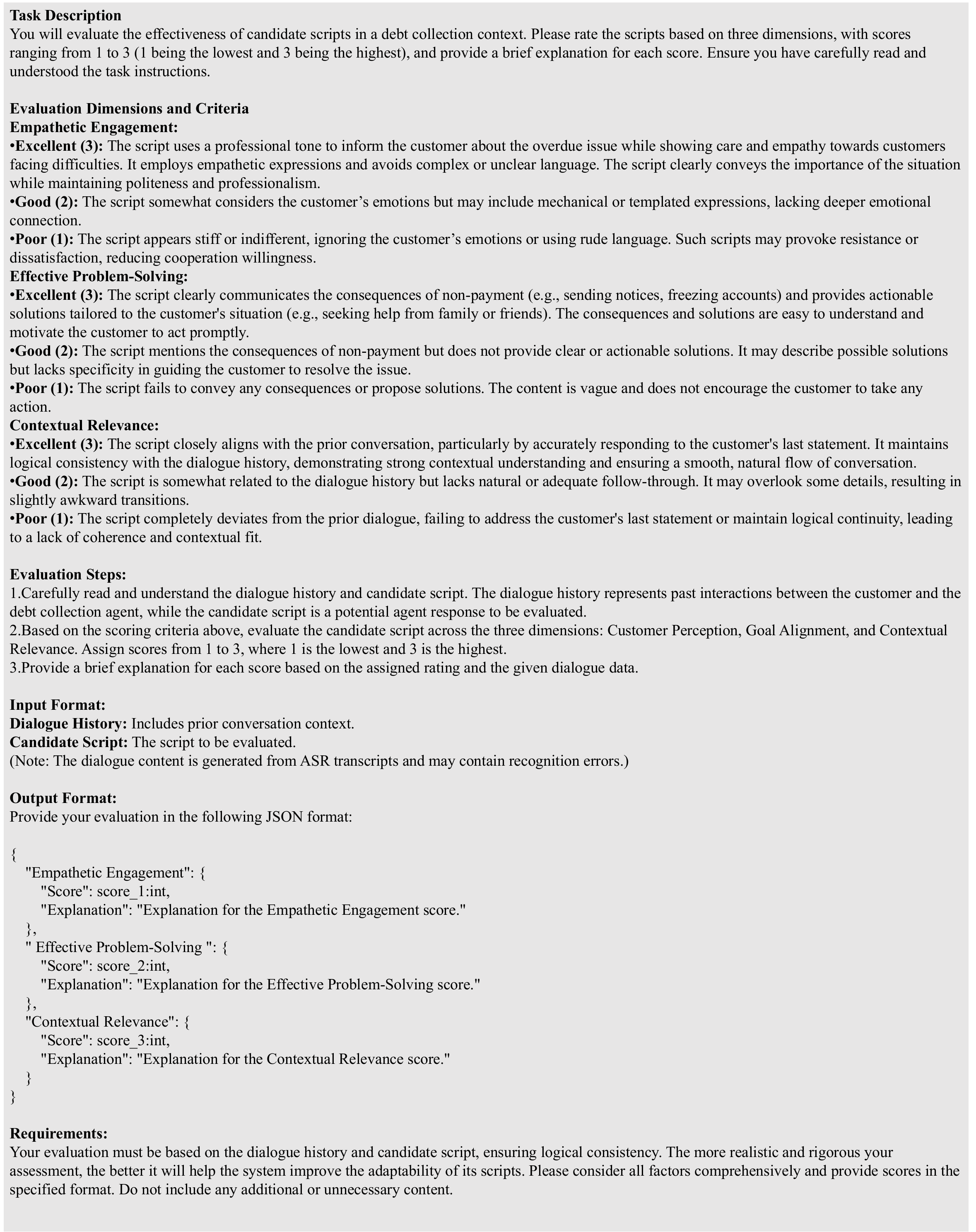} 
    \caption{The prompt used for LLM evaluation.}
    \label{fig:prompt}
\end{figure*}

\end{document}